\documentstyle[12pt,epsfig,amssymb]{article}
\def\Tr{{\rm Tr}}
\def\d{\partial}
\newcommand{\eq}[1]{(\ref{#1})}

\begin{document}
\title{Trends in Grand Unification: unification at strong coupling and
composite models\footnote{Lecture 
at the Baksan School ``Particles and Cosmology'', Baksan Valley, April
19 - 25, 1999.}}
\author{V.A.Rubakov and S.V.Troitsky\\
{\small\em
Institute for Nuclear Research of the Russian Academy of Sciences,}\\
{\small\em 60th October Anniversary Prospect 7a, Moscow 117312 Russia.}
}
\date{}
\maketitle
\vspace{-12mm}
\begin{abstract}
{We review several problems of conventional Grand
Unification and some new approaches. In particular, we discuss
strongly coupled Grand Unified Theories.  Standard Model may emerge as a low
energy effective theory of composite particles in these models.  We
construct a realistic model of this kind. }
\end{abstract}

\section{Conventional unification.}
\label{Reasons}
While the Standard Model of elementary particles provides a good
description of Nature at energies accessible to current experiments,
it is widely believed that the variety of interactions operating in
very different ways at our energy scales originate from a single
fundamental interaction at high energies. The standard lore (for a
review see Ref.~\cite{GUT}) is that (a)~the strong, weak, and
electromagnetic interactions unify at energy scale $M_{\rm GUT}\sim
10^{16}$~GeV; (b)~the Standard Model is supersymmetric at energies
above $M_{\rm SUSY}\sim 1$~TeV; (c)~there are no new particles
participating in gauge interactions with masses between $M_{\rm SUSY}$ and $M_{\rm
GUT}$ (``the Grand Desert''). This logical possibility is supported by
the unification of coupling constants.  Running gauge couplings of the
Minimal Supersymmetric Standard Model (MSSM) exhibit approximate
unification at high energy scale $M_{\rm GUT}$ if their evolution is
described by perturbative one- or two-loop renormalization group (RG)
equations without extra charged matter heavier than $M_{\rm
SUSY}$. The same is true, though parameter-dependent ($\tan\beta$),
for some of the Yukawa couplings ($b-\tau$ unification). The
assumption of the Grand Desert is important for this conclusion.
Supersymmetry is usually introduced because it helps to solve the
problem of stability of the Higgs mass in the Standard Model against
the radiative corrections.  It is also considered as a necessary
ingredient of Grand Unification for two main reasons. First, it
results in much better unification of couplings. Second, the
unification scale is about two orders of magnitude higher than that in
theories without supersymmetry. The latter fact is important for
better consistency with the observed absence of the proton decay.

However, almost every particular realization of the conventional Grand
Unification scenario meets considerable problems. Let us outline
briefly a few of them.

{\em Doublet-triplet splitting.} In most popular versions of 
unification, weak $SU(2)$ and colour $SU(3)$ gauge groups are subgroups
of a simple group, and all matter fields should fall into
complete multiplets of this unification group. As an example, weakly interacting
leptons belong to one and the same GUT multiplet as coloured
quarks. In a similar way, scalar
Higgs doublets ought to have their coloured counterparts. These strongly
interacting particles (triplets in the simplest $SU(5)$ model) mediate
proton decay at unacceptable rate, unless they are superheavy, i.e.\
have 
masses of order $M_{\rm GUT}$ or higher. Unnatural mass difference between
particles of one and the same multiplet, doublet and triplet Higgses,
can be cured either by fine tuning of parameters or by choosing very
complicated Higgs sector.

{\em Potential phenomenological problems.} With growing experimental
accuracy, it appears that GUTs predict a bit too fast proton decay
\cite{proton}, and also that the coupling constants do not quite merge
\cite{Langacker} (unification requires $\alpha_s(M_Z)$ slightly larger than measured).

{\em Supersymmetry breaking.} If supersymmetry has anything to do
with reality, it has to be broken in the low energy theory.
A mechanism that breaks supersymmetry in a
phenomenologically acceptable way in the (supersymmetric extensions of
the) Standard Model is not known; therefore, another, completely different
sector is usually introduced. Supersymmetry breaking in this new,
``hidden'' or ``secluded'' sector is often associated with strong
interactions due to new gauge fields. These new gauge interactions
are not unified with our interactions in conventional
GUTs. So, one started from the idea of unification but arrived at
a separate gauge interaction in the supersymmetry breaking sector.

{\em Grand Desert.} Experimental data and their theoretical analysis
point towards the necessity of several mass scales in the Grand
Desert. Some of the arguments in favour of the new scales are:
\begin{itemize}
\item
The most popular solution to the 
{\em strong $CP$ problem} requires the axion scale of order
$10^{10}$~GeV
(see, e.g., Ref.~\cite{axion}).
\item
{\em Non-vanishing neutrino masses}, in case they are provided by the see-saw or
similar mechanism, point towards the mass scale of order $10^{12}$ --
$10^{14}$~GeV (mass of right-handed neutrino)
(see, e.g., Ref.~\cite{neutrino}).
\item
{\em Models of gauge mediation of supersymmetry breaking} make use of the
mass scale of messenger fields of order $10^8$ -- $10^{14}$~GeV
(see, e.g., Ref.~\cite{GMSB}).
\end{itemize}
All these arguments suggest that the Grand Desert is actually
populated with new particles of various masses. 

\section{New trends}
\label{new}
Is it possible to relax the assumption about the Grand Desert and, at the
same time, preserve the self-consistent picture of unification? When a
few matter fields are added with masses between $M_{\rm SUSY}$ and
$M_{\rm GUT}$, the gauge coupling unification is preserved at the one-loop
level, provided the new states fall into complete representations of a
GUT gauge group.  However, the two-loop analysis (which is adequate in
view of the current experimental accuracy) shows that the unification of
couplings becomes considerably worse, as compared to the theory
with Grand Desert \cite{CaroneMurayama}.

On the other hand, with new matter states added, the beta functions
change in such a way that the gauge couplings become larger in the
ultraviolet. Already with a few additional fields, the asymptotic
freedom of QCD is lost, and perturbative RG analysis at high energies
may not be applicable. Indeed, the one-loop RG equations for MSSM
gauge couplings are
\begin{equation}
{d\alpha_i\over dt}=-b_i\alpha_i^2,
\label{RG}
\end{equation}
where $\alpha_i$, $i=1,2,3$, denote the gauge couplings of $U(1)$,
$SU(2)$ and $SU(3)_C$
gauge groups, respectively, the first coefficients of the beta functions are
$b_1=-33/5$, $b_2=-1$, $b_3=3$, $t={1\over 2\pi}\ln{Q\over M_{\rm GUT}}$ and
$Q$ is the momentum scale. The solution to Eqs.~\eq{RG} is
$$
\alpha_i^{-1}(Q)=\alpha_i(M_{\rm GUT})^{-1}+b_i t.
$$
For $b_i<0$, the corresponding couplings grow at high energies.
 When new matter fields are added, $b_i$
decrease. Suppose that additional particles fall into complete
vector-like multiplets of, say, $SU(5)$ unified gauge group, for
example, $n_5$ of ($5+\bar{5}$) or $n_{10}$ of ($10+\overline{10}$).  Then
the coefficients $b_i$ are shifted uniformly, 
$$b'_i=b_i-n,$$ 
where
$n=n_5+3n_{10}$. For $n>3$, all three gauge groups are not
asymptotically free, and their coupling constants grow in the
ultraviolet. For $n\ge 5$, the coupling constants reach their Landau
poles below the Planck scale. With measured low energy values of
$\alpha_i$ and appropriately chosen masses of additional matter
states, however, the three Landau poles coincide (at the one loop, this
common Landau pole is always located at $M_{\rm GUT}\sim 10^{16}$~GeV). This
phenomenon is called strong unification (see Fig.~1).
\begin{figure}
\begin{picture}(0,160)%
\centerline{\epsfxsize=1.0\textwidth \epsfbox{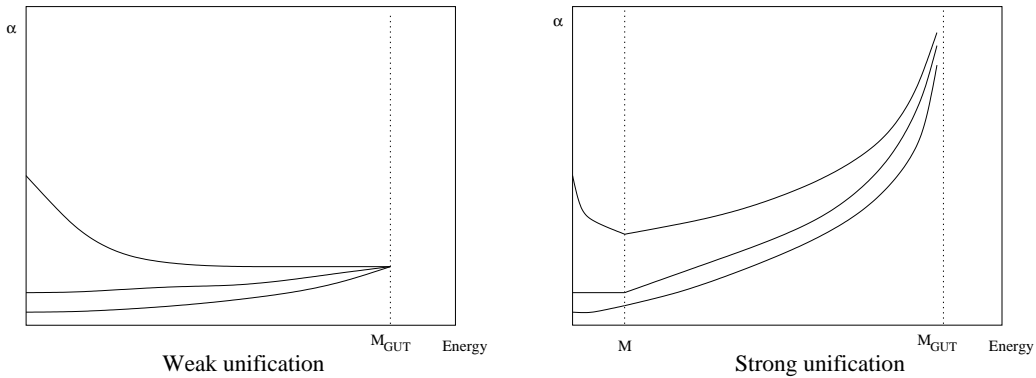}}
\end{picture}%
\caption{Sketch of running of gauge coupling constants in weak and
strong unification scenarios.}
\end{figure}

Of course, growing gauge couplings at high energies do
not contradict low energy data (in particular, observed asymptotic
freedom of QCD) because new matter fields affect RG
evolution only at energies of order of their masses and higher.

The possibility that unification may occur in the non-perturbative
domain was
considered long ago \cite{STrongUnif} and clearly relaxes the
unification constraints because non-perturbative evolution of couplings
is unknown. Despite the latter fact, these models are highly
predictive because the low energy evolution of gauge couplings is
governed by strongly attractive infrared (IR) fixed points \cite{Ross}. The RG
equations \eq{RG}, written in terms of the {\em ratios} of coupling
constants,    
$$
{d\over dt}\ln{\alpha_j\over\alpha_i}=b'_i\alpha_i-b'_j\alpha_j,
$$
have
infrared fixed points,
$$
{\alpha_i\over\alpha_j}={b'_j\over b'_i},
$$
which are stable for $b'_i<0$.
In practice, they are so strongly attractive that
even at $n=5$, the ratios are almost constant at
$Q<0.04 M_{\rm GUT}$. This means that the low energy behaviour of the
couplings is independent of the details of their non-perturbative
evolution near $M_{\rm GUT}$, thus keeping the theory predictive.

On the contrary, dynamics at very high energies, of order $M_{\rm GUT}$ and
higher, cannot be controlled in the usual way since the theory is
strongly coupled at these scales. Various models of high-energy theory
can be suggested.  Nature may be described at very high energies by a
string theory where non-perturbative couplings appear quite
naturally. 
Alternatively, the fundamental theory above
$M_{\rm GUT}$ may be some unified gauge theory which is asymptotically
free and has confinement scale of order
 $M_{\rm GUT}$ (another possibility is that this theory is
approximately scale-invariant ``in the infrared'').  

One way to guess the dynamics operating at energies higher than
$M_{\rm GUT}$ in strong unification scenario is to consider the
Standard Model as a low energy description of a more fundamental
strongly coupled theory, in the same way as the sigma model provides
the low energy description of QCD. If MSSM is a low energy effective
theory, the low energy degrees of freedom, i.e., quarks, leptons, and
gauge and Higgs bosons, are composite particles. The compositeness of
all these particles used to be problematic because in most models,
composite fermions had masses of the order of the compositeness scale
(baryons are as heavy as $\Lambda_{\rm QCD}$), and no mechanism
leading to composite massless gauge bosons was known (in the QCD case,
there are no light vector bosons; $m_\rho\sim \Lambda_{\rm QCD}
$). These two properties were major obstacles to construct composite
version of the Standard Model with its gauge bosons and
chiral fermions which are almost massless compared to possible
compositeness scale (which is not smaller than a few TeV according to
current experimental bounds \cite{PDG}).

Remarkably, both problems are in principle solved by supersymmetry.  In the last
few years, outstanding progress has been made in understanding the
dynamics of strongly coupled supersymmetric gauge theories (see
Ref.~\cite{PSV} for reviews). In particular, supersymmetry is so
restrictive that sometimes a weakly coupled theory may be uniquely
inferred which describes the infrared dynamics of a model with
strong coupling at long distances (like in the QCD case). However,
unlike QCD, the low energy effective theory of composite particles
{\em is} in a number of cases a theory with massless fermions and
massless composite gauge bosons. This phenomenon is known as $N=1$
duality, and the high- and low-energy theories are often called dual
theories.  The famous (and first) example \cite{Seiberg} is
supersymmetric QCD (supersymmetric $SU(N_c)$ gauge theory with $N_f$
flavours of ``quark'' supermultiplets in the fundamental and
antifundamental representations of the gauge group). At
$N_c+1<N_f<{3\over 2}N_c$, the gauge group is strongly coupled in the
infrared while the effective low energy theory of composite particles is
$SU(N_c-N_f$) gauge theory with massless matter supermultiplets that
carry non-trivial quantum numbers under the low energy gauge
group. Many more examples are known (some of them are reviewed in
Ref.~\cite{PSV}); their common feature is that the fundamental theory
is rather simple but its low energy counterpart may be quite complex
and may contain product groups with matter multiplets in chiral
representations, very similar to the (supersymmetric version of the)
Standard Model.

\section{Examples of strongly coupled GUTs}
\label{ex}
Construction of a realistic composite Standard Model is a non-trivial
task; the solution to this problem is not unique and requires 
guesswork. One possibility is based on the conventional picture of
Grand Unification in the weak coupling regime. The Grand Unified
Theory itself is then an effective theory of composite particles.  An
example of a model of this kind was suggested in Ref.\
\cite{Strassler}. That model is based on the fundamental gauge group
$SU(N)\times Sp(4)\times Sp(6)$ with $N\ge 17$ (!); its low energy
description is provided by $SO(10)$ GUT, broken down to the Standard
Model by means of the usual Higgs mechanism at weak coupling. Another
possibility, which does not require enormous gauge groups of the
fundamental theory,
is to invoke the idea of strong unification. We present below a class
of models with simple fundamental gauge group and MSSM with
extra vector-like matter as its low energy description. These models
are strongly coupled at GUT scale but, in some sense, represent
analytic continuations of usual weakly coupled GUTs.

\subsection{The class of models.}
Consider an asymptotically free gauge theory with gauge group $G$ and
a certain moduli space. At some submanifold of the moduli space, the
group $G$ is broken down to its subgroup $G_L\subset G$. Choose matter
content in such a way that $G_L$ is IR free. Then the low energy
theory at these points of moduli space is described by $G_L$. It is
often possible to add a tree level superpotential that singles out this
specific vacuum.

Let the superpotential have a minimum where the expectation values of
matter fields are of order $v$ and $G$ is broken down to $G_L$. Let
$\Lambda$ be the scale at which $G$ is strongly coupled. At
$v\gg\Lambda$, the theory is weakly coupled, and the usual Higgs
mechanism is in operation. Thus, at energies below $v$, in particular,
in the IR limit, the model is described by $G_L$ gauge theory for any
$v\gg\Lambda$. Let us change smoothly the parameters of the
superpotential in such a way that $v$ becomes smaller than
$\Lambda$. Some Green's functions, in particular, those indicating
which gauge group remains unbroken, exhibit holomorphic dependence of
the parameters. Hence, even at $v\lesssim\Lambda$, the low energy
effective theory is again the same $G_L$ gauge theory, free in the
IR. This argument enables one to establish the low energy effective
description even though at intermediate energies of order $\Lambda$
the model is strongly coupled, and detailed description of dynamics,
in particular, of symmetry breaking, is impossible. We will choose
$G$, matter content, and superpotential in such a way that the low
energy theory ($G_L$) is just the MSSM with extra vector-like
matter. Clearly, the model will be an analytical continuation of a
usual weakly coupled GUT from $v\gg\Lambda$ down to $v\lesssim\Lambda$.

Very similar arguments appear in the discussion of duality in $SU(n)$
gauge theories with adjoint chiral superfield $\Phi$ and a
superpotential, Ref.\ \cite{Kutasov}. Without superpotential, the moduli
space is described by gauge invariants made of powers of $\Phi$,
namely $\Tr\Phi^2$, \dots, $\Tr\Phi^{n-1}$. At a generic point of the
moduli space, $SU(n)$ is broken down to $U(1)^{n-1}$. There are, however,
some special points of extended symmetry where $G_L=SU(k)\times
SU(n-k)\times U(1)$. These points remain supersymmetric vacua if the
superpotential,
$$
W=m\Tr\Phi^2+\lambda\Tr\Phi^3
$$
is added. Consider
the case in which, in addition to $\Phi$, there are $p$ fundamental
flavours, $Q$, $\bar Q$. Then the first coefficient of the beta
function of $G$ is $b_0^{(n)}=3n-p$, whereas for subgroups of $G_L$ one
has $b_0^{(n-k)}=3(n-k)-p$, $b_0^{(k)}=3k-p$. The number $p$ can be
adjusted in such a way that $G$ is asymptotically free while $G_L$ is
free in the infrared, so the low energy theory is described by
$G_L$. At intermediate energies, the theory is in the conformal phase,
and two alternative descriptions are possible --- one in terms of
(strongly coupled) $G_L$, another in terms of its dual, $SU(p-k)\times
SU(p-n+k)\times U(1)$; both theories are in a strongly coupled phase,
and detailed description of the dynamics is not possible.

\subsection{An example.}
Consider now the implementation of these arguments to the case where
$G_L=SU(3)\times SU(2)\times U(1)$.  The first coefficient of beta
function of $G$ is equal to $b_0^{(G)}=3l-h-7-e$, where $3l$ is the contribution
of the gauge superfield, $h$ is the contribution of heavy matter
superfields, and $7+e$ corresponds to light superfields. Here, 7 comes
from the MSSM matter (3 chiral generations and electroweak Higgses)
and $e$ is the contribution of extra vector-like matter fields. Since
$G$ should be asymptotically free, one needs $b_0^{(G)}>0$.  The most
restrictive subgroup of $G_L$ is $SU(3)$ for which $b_0^{(SU(3))}=3\cdot
3-6-e$ (6 instead of 7 the since Higgs triplets are supposed to be
heavy). It is known \cite{Ross} that for unification at strong
coupling below Planck mass it is required that $b_0^{(SU(3))} <-1$. So, one has
rather strong restriction on the number of extra matter states,
\begin{equation}
4<e<3l-h-7.
\label{ineq}
\end{equation}
For the simplest $SU(5)$ GUT, $l=h=5$, so the inequality \eq{ineq}
cannot be satisfied. We will discuss $SO(10)$ case in what follows.

The contributions of the lowest $SO(10)$ multiplets to $b_0$ are:
\begin{center}
\begin{tabular}{rcl}
vector {\bf 10} & $\mapsto$ & 1\\
spinor {\bf 16} & $\mapsto$ & 2\\
adjoint {\bf 45} & $\mapsto$ & 8\\
symmetric tensor {\bf 54} & $\mapsto$ & 12.
\end{tabular}
\end{center}
$SO(10)$ can be broken down to $G_L$ in two different ways. The
most popular way ivolves Pati--Salam group at the intermediate stage,
and requires heavy {\bf 54} or higher representations. With $h=12$,
however, the inequality \eq{ineq} cannot be satisfied. Another way is
to embed $G_L\subset SU(5) \subset SO(10)$. This breaking may be
achieved by means of (a) heavy adjoint field (see Appendix) or (b)
heavy adjoint and ${\bf 16}+\overline{\bf 16}$. In the case (a), one
needs non-renormalizable superpotential and $G_L$ is not just the MSSM
gauge group but contains extra $U(1)'$ factor; one has $4<e<9$ in that
case. In the case (b), 
the inequality \eq{ineq} may be satisfied only if ${\bf
16}+\overline{\bf 16}$ can be arranged to be light and contribute to
$e$ rather than $h$. 

The theory has two essential mass scales: $\Lambda$, where $SO(10)$
becomes strongly coupled, and $v\sim\sqrt{mM}$, where $\Phi$
condenses\footnote{One has to keep in mind that at strong coupling,
only gauge invariant degrees of freedom make sense, and vacuum
expectation value of $\Phi$, Eq.~\eq{vev}, is to be understood in terms
of the expectation values of $\Tr\Phi^2$,\dots. We keep weak-coupling
notation for simplicity.} (see Appendix). In the ultraviolet, at energies
$\gg\Lambda$, the theory is weakly coupled and correctly described by
$SO(10)$. In the infrared, at energies $\ll v$, correct weakly coupled
description is provided by $G_L$ (at very low energies, effects of
supersymmetry, electroweak symmetry and $U(1)'$ symmetry breaking
should be taken into account as well as decoupling of extra
matter). If $v\gg\Lambda$, this is a usual weakly coupled GUT;
however, at $v\lesssim\Lambda$ the theory is strongly coupled at
intermediate energies. In fact, it is expected that the theory is in
the conformal phase there, and one of its descriptions is provided
(in the case when heavy vector-like matter is in {\bf 10}s) by the
dual theory of Ref.~\cite{Strassler}. The
sketch of running of coupling constants is presented in Fig.~2. 

\begin{figure}
\begin{picture}(0,160)%
\centerline{\epsfxsize=1.0\textwidth \epsfbox{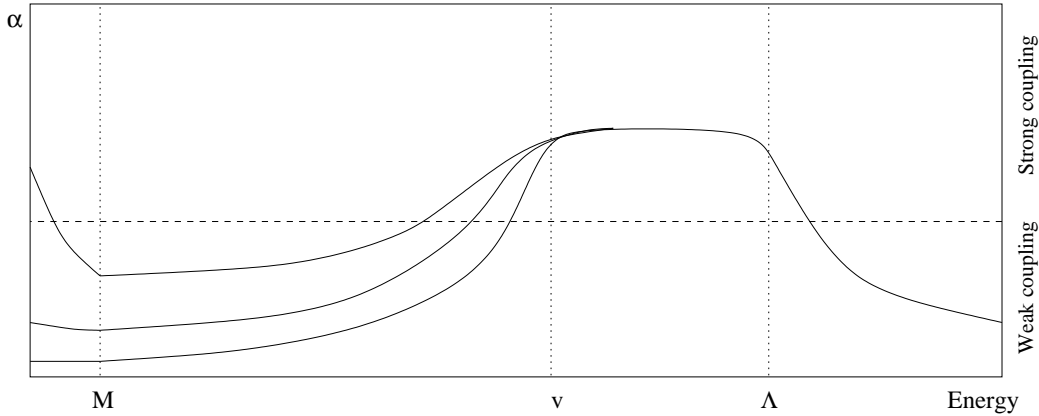}}
\end{picture}%
\caption{Sketch of running of gauge coupling constants in $SO(10)$ model.}
\end{figure}

Further possibilities emerge in the case of $E_6$ gauge group where
additional vector-like matter states are necessary counterparts of
MSSM matter since the lowest $E_6$ representation, the fundamental, is
decomposed under $E_6\to SO(10)\times U(1)''$ as ${\bf 27}\to {\bf 16}
+ {\bf 10} + {\bf 1}$. 

Since the models of the class outlined here are merely analytically
continued weakly coupled GUTs, many of usual problems remain
unsolved. In particular, the proton lifetime bound imposes usual restrictions on the scale $v$ which is a characteristic
mass scale of vector leptoquarks. One might hope that in the conformal
phase, usual calculation of proton decay width is invalid since no
asymptotic states like leptoquarks are present and the only important scale is
$\Lambda$ (in that case, an interesting possibility of low energy
unification would be allowed). However, this is not the case since
Green's functions involving baryon number violating operators may still
be estimated, and corresponding
amplitudes will be suppressed in the usual way by the mass of virtual
leptoquarks. 

It is assumed, like in usual weakly coupled GUTs, that supersymmetry
breaking is provided by different mechanisms. $U(1)'$ is broken 
if ${\bf 16}+\overline{\bf 16}$ are involved besides the
adjoint; if only adjoint breaks $SO(10)$ (with non-renormalizable
superpotential) then $U(1)'$ should also be broken by a different
mechanism.

\section{Outlook}
\label{out}
Strong unification and compositeness may be useful for
resolving some of the long-standing difficulties of unification. A
low-energy theory of composites does not necessarily contain Higgs
triplets since there is no reason to have complete GUT multiplets in
the effective theory. Proton could be protected from decay by
non-anomalous global symmetries of the theory (which cannot be just
baryon number because the latter had to be broken in the early
Universe). The scale of GUT symmetry breaking may be relatively low in
this case. At the same scale (which corresponds to the strongly
coupled phase), supersymmetry can be broken dynamically in a
phenomenologically acceptable way, thus avoiding the hidden
sector. 

These problems could be solved if the MSSM with extra vector-like
matter is a dual theory of a strongly coupled GUT, so that $G_L$ is
not a subgroup of $G$ and no fundamental superpotential is involved,
like in the case of SQCD \cite{Seiberg}.  Clearly, much work
has to be done before a realistic model is
constructed. Phenomenological and cosmological implications of strong
unification scenarios of this kind deserve further investigation.

The authors are indebted to S.L.Dubovsky for numerous helpful
discussions. This work is supported in part by RFBR grant 99-02-18410.
Work of S.T.\ is supported in part by the Russian Academy of Sciences
Junior Researcher Project No. 37.

\appendix
\section{Symmetry breaking.}
The most general matrix from $so(10)$ algebra is
$$
\left(
\begin{array}{cc}
A_1 & B \\
~\\
-B^T & A_2
\end{array}
\right).
$$
where $A_i$, $B$ are $5\times 5$ matrices and $A_{1,2}^T=-A_{1,2}$.
If $A_1=A_2$, $B=B^T$, $\Tr B=0$, then matrices $(A+iB)$ form the
standard $su(5)$ algebra (with anti-hermitian generators). The adjoint
field $\Phi$ belongs to $so(10)$, too, and its vacuum expectation
value with $A_{1,2}^\Phi=0$, $B^\Phi={\rm diag}(a,a,a,b,b)$ breaks
$SO(10)$ down to $G_L=SU(3)\times SU(2)\times U(1)\times U(1)'$ where
the Standard Model group is embedded in $SU(5)$ in the usual way, and
$U(1)'$ generator has nonzero $\Tr B$, for example,
$$
\left(
\begin{array}{cc}
0 & 1_{5\times 5} \\
~\\
-1_{5\times 5} & 0
\end{array}
\right).
$$
Symmetry breaking by adjoint vev is described by the following simple
rule \cite{Slansky}. Take the Dynkin diagram for the full symmetry
group. Remove one dot and add $U(1)$ factor instead of it. Different 
subgroups which may be obtained after breaking with adjoint field
correspond to different dots removed. To remove one more dot one needs
either the second adjoint or non-renormalizable superpotential. In our
case, removing one dot from
$$
\begin{array}{c}
\begin{picture}(95,40)(0,-20)
\put(0,0){\line(1,0){60}}
\put(60,0){\line(2,1){30}}
\put(60,0){\line(2,-1){30}}
\multiput(0,0)(30,0){3}{\circle*{5}}
\put(90,15){\circle*{5}}
\put(90,-15){\circle*{5}}
\end{picture}
\\
SO(10)
\end{array}
$$
results in
$$
\left(
\begin{array}{c}
\begin{picture}(95,10)(0,-5)
\put(0,0){\line(1,0){90}}
\multiput(0,0)(30,0){4}{\circle*{5}}
\end{picture}
\\
SU(5)
\end{array}
\right)
\times U(1)
~~~~
{\rm or}
~~~~
\left(
\begin{array}{c}
\begin{picture}(65,40)(0,-20)
\put(0,0){\line(1,0){30}}
\put(30,0){\line(2,1){30}}
\put(30,0){\line(2,-1){30}}
\multiput(0,0)(30,0){2}{\circle*{5}}
\put(60,15){\circle*{5}}
\put(60,-15){\circle*{5}}
\end{picture}
\\
SO(8)
\end{array}
\right)
\times U(1)
$$
$$
{\rm or}
~~~~
\left(
\begin{array}{c}
\begin{picture}(95,10)(0,-5)
\put(30,0){\line(1,0){60}}
\multiput(0,0)(30,0){4}{\circle*{5}}
\end{picture}
\\
SU(2)\times SU(4)
\end{array}
\right)
\times U(1)
~~~~
{\rm or}
~~~~
\left(
\begin{array}{c}
\begin{picture}(95,10)(0,-5)
\put(0,0){\line(1,0){30}}
\multiput(0,0)(30,0){4}{\circle*{5}}
\end{picture}
\\
SU(3)\times SU(2)\times SU(2)
\end{array}
\right)
\times U(1).
$$
Thus, to obtain
$$
\left(
\begin{array}{c}
\begin{picture}(65,10)(0,-5)
\put(0,0){\line(1,0){30}}
\multiput(0,0)(30,0){3}{\circle*{5}}
\end{picture}
\\
SU(3)\times SU(2)
\end{array}
\right)
\times U(1) \times U(1)'
$$
two dots should be removed. Since introducing the second heavy adjoint
contradicts Eq.~\eq{ineq}, non-renormalizable superpotential is
required. If
$$
W=m\Tr\Phi^2-{1\over M}\Tr\Phi^4,
$$
then the equations for supersymmetric minimum written in terms of
$A^\Phi$, $B^\Phi$,
$$
{\d W\over \d A^\Phi}={\cal{O}}(A^\Phi),
$$
$$
{\d W\over \d B^\Phi}={\cal{O}}(A^\Phi)+4mB^\Phi -{6\over M}(B^\Phi)^3,
$$
have a solution with 
\begin{equation}
A^\Phi=0,~~~~B^\Phi=\sqrt{2mM\over 3}\,
{\rm diag}(1,1,1,-1,-1).
\label{vev}
\end{equation}
This vacuum corresponds to unbroken $G_L=SU(3)\times SU(2) \times U(1)
\times U(1)'$.

\def\pl#1#2#3{{\it Phys.\ Lett.} {\bf B#1} (19#2) #3}
\def\pr#1#2#3{{\it Phys.\ Rev.} {\bf D#1~} (19#2) #3}
\def\np#1#2#3{{\it Nucl.\ Phys.} {\bf B#1~} (19#2) #3}

\end{document}